# Deep learning-based air temperature mapping by fusing remote sensing, station, simulation and socioeconomic data

Huanfeng Shen[a,d,e], Yun Jiang[a], Tongwen Li[a], Qing Cheng[b], Chao Zeng[a,*], Liangpei Zhang[c,d]

[a] School of Resource and Environmental Sciences, Wuhan University, Wuhan 430079, China.

[b] School of Urban Design, Wuhan University, Wuhan 430079, China.

[c] The State Key Laboratory of Information Engineering in Surveying, Mapping and Remote Sensing, Wuhan University, Wuhan 430079, China.

[d] Collaborative Innovation Center of Geospatial Technology, Wuhan 430079, China.

[e] The Key Laboratory of Geographic Information System, Ministry of Education, Wuhan University, Wuhan 430079, China.

[*] Corresponding author: Chao Zeng (zengchaozc@hotmail.com)

## ABSTRACT

Air temperature (Ta) is an essential climatological component that controls and influences various earth surface processes. In this study, we make the first attempt to employ deep learning for Ta mapping mainly based on space remote sensing and ground station observations. Considering that Ta varies greatly in space and time and is sensitive to many factors, assimilation data and socioeconomic data are also included for a multi-source data fusion based estimation. Specifically, a 5-layers structured deep belief network (DBN) is employed to better capture the complicated and non-linear relationships between Ta and different predictor variables. Layer-wise pre-training process for essential features extraction and fine-tuning process for weight parameters optimization ensure the robust prediction of Ta spatio-temporal distribution. The DBN model was implemented for 0.01° daily maximum Ta mapping across China. The ten-fold cross-validation results indicate that the DBN model achieves promising results with the RMSE of 1.996°C, MAE of 1.539°C, and R of 0.986 at the national scale. Compared with multiple linear regression (MLR), back-propagation neural network (BPNN) and random forest (RF) method, the DBN model reduces the MAE values by 1.340°C, 0.387°C and 0.222°C, respectively. Further analysis on spatial distribution and temporal tendency of prediction errors both validate the great potentials of DBN in Ta estimation.

**1. Introduction**

Air temperature (Ta) is one of the fundamental meteorological parameters and has been associated with a wide range of studies including disease vectors propagating and human health (Li et al., 2010; Lowen et al., 2007), terrestrial hydrology and phenology (Lin et al., 2012; Wang et al., 2009), climate and environment change (Robeson, 2002). Typically, Ta is measured through monitoring stations at 2 m above the ground with high precision. However, the spatial distribution of the meteorological stations may be extremely sparse on a large scale, especially in some underdeveloped and complicated terrain areas. Hence, traditional spatial interpolation methods, such as Kriging, Inverse Density Weighting (IDW) and Spline interpolation have been used to generate spatially continuous Ta. However, these interpolation methods are still limited to the station density and the complexity of different environmental conditions (Shi et al., 2017; Ung et al., 2001).

Related researches have confirmed that there exists an energy exchange between land surface and near-surface atmosphere, land surface temperature (LST) retrieved from the thermal infrared remote sensing truly has a strong physical relationship with the Ta. Recently, the satellite-derived LST products with high temporal and spatial resolution is widely applied to estimate Ta (Colombi et al., 2007; J. Stoll and Brazel, 2013; Lin et al., 2012; Tomlinson et al., 2012; Zakšek and Schroedter-Homscheidt, 2009). Nevertheless, LST cannot be directly regarded as a proxy for Ta in terms of their different physical meaning and magnitude (Jin and Dickinson, 2010), and the LST-Ta relationship is sensitive to many spatio-temporal factors in reality, especially for the maximum Ta (Jin et al., 1997; Lu et al., 2018). How to accurately estimate Ta with a large spatial distribution has become one of the research hotspots in the field of remote sensing. Various satellite-based parameterization algorithms have been implemented to estimate Ta and can be divided into three main types (Ho et al., 2014; Zhu et al., 2017).

The first type is the temperature-vegetation index (TVX) method, which is a spatial method based on the presumption that vegetation canopy temperature approximates near-surface Ta in an absolutely thick canopy (Nieto et al., 2011; Prihodko and Goward, 1997; Zhang et al., 2014). The strong negative correlation between LST and normalized difference vegetation index (NDVI) is adopted to extrapolate the Ta. According to Xu et al. (2011), some additional rules had been made for the TVX method to broaden the applied range, and the results demonstrated good accuracy and applicability in cropland areas in crop growing seasons. In another study, Zhu et al. (2013) improved the accuracy of daily maximum Ta estimation by using the TVX method. Although TVX method performed adequately in some studies, the basic presumption makes it unfeasible to estimate Ta for regions or seasons without high vegetation cover (Vancutsem et al., 2010; Zhu and Zhang, 2011).

The second type is the energy balance method that has a clear physical mechanism. The net radiation is deemed to be equal to the sum of the surface's sensible, soil and latent heat fluxes (Meteotest, 2010; Zhang et al., 2015). According to the energy balance equation, Ta can be linked to the LST and other surface environmental parameters. Sun et al. (2005) once presented a thermodynamics-based method to derive the relationship between Ta and LST by using aerodynamic resistance and a crop water stress index obtained from Moderate Resolution Imaging Spectroradiometer (MODIS) data. In the work conducted by Hou et al. (2013), an Energy Balance Bowen Ratio model was developed with the mean retrieval error of approximately 2.21°C. This method shows good portability and universality, however, it may need comprehensive parameters as inputs, which are hard to obtain directly (Mostovoy et al., 2006).

The third and most commonly-used type is the statistical method, which is typically based on the regressive relationship between Ta and other variables. Simple statistical model only structures a linear regression between LST and Ta (Shi et al., 2017; Vogt et al., 1997), while advanced statistical models, such as multiple regression model, artificial neural network and machine learning models use a mass of auxiliary variables to establish linear or non-linear relationship (Fu et al., 2011; Mohsenzadeh Karimi et al., 2018; Singh et al., 2006). For instance, Janatian et al. (2017) proposed an advanced statistical framework by constructing fourteen statistical models through a stepwise regression analysis based on MODIS LST data and other variables. The geographically weighted regression (GWR) model, as a widely used statistical approach, was once used and confirmed to have better performance than ordinary linear regression (OLS) model (Yao and Zhang, 2013). Besides, daily GWR models were developed to produce daily Ta for urban and surrounding areas in the conterminous United States recently (Li et al., 2018). The increasing popularity of machine learning stemmed in Ta estimation field in recent years. To date, various machine learning models have been developed and reported successfully in Ta estimation. For example, Li and Zha (2018) applied the random forest model (RF) to estimate relative humidity and temperatures in hot summer over China and achieved acceptable prediction errors. Noi et al. (2017) estimated daily Ta from dynamic combinations of MODIS LST data by comparing multiple linear regression (MLR), cubist regression (CB), and RF. Besides, Xu et al. (2018) even implemented and compared ten machine learning algorithms to estimate monthly Ta in the Tibetan Plateau and the results showed that machine learning algorithms had great potentials in Ta estimation.

As mentioned above, previous studies have confirmed that machine learning algorithms have great advantages in Ta estimation due to the capacity of handling non-linear relationships. Deep learning, as a well-known new generation of technology in machine learning methods, has been proven to be very promising in many domains of researches (Kuwata and Shibasaki, 2015; Shen et al., 2018; Song et al., 2016). However, to the best of our knowledge, deep learning has never been used for Ta estimation. Hence, it is of great interest to examine whether deep learning technique show more advantages for this complicated non-linear problem in Ta estimation. For this, a 5-layers deep belief network (DBN) is structured to establish the relationship between station Ta observation and multi-source data including remotely sensed data, socioeconomic data and assimilation data. The network model is then used for high spatio-temporal resolution Ta mapping across China. Its effectiveness and advantages are validated by extensive experiments.

**2. Study area and data**

## 2.1. Study area

China is selected as the study area in this research (Fig. 1). The total area covers approximately 9.6 million km$^2$, lies between latitudes 3°N and 54°N, and between longitudes 73°E and 136°E. Additionally, China shows highly spatial heterogeneity of land-cover types, the plains and basins account for about 33% of the land area, while mountainous, hills and plateaus account for about 67%. According to the geographical division of China, there are seven zones including eastern, northern, southern, central, southwestern, northwestern and northeastern China. In general, the terrain is high in the southwest but low in the east with the whole elevation ranging from -154 m to 8848 m. The southwest of China has the tallest Tibetan Plateau in the world, with an average elevation of more than 4000 meters. Due to the highly complex terrain, the climate in China varies greatly in space and is mainly dominated by dry seasons and wet monsoon.

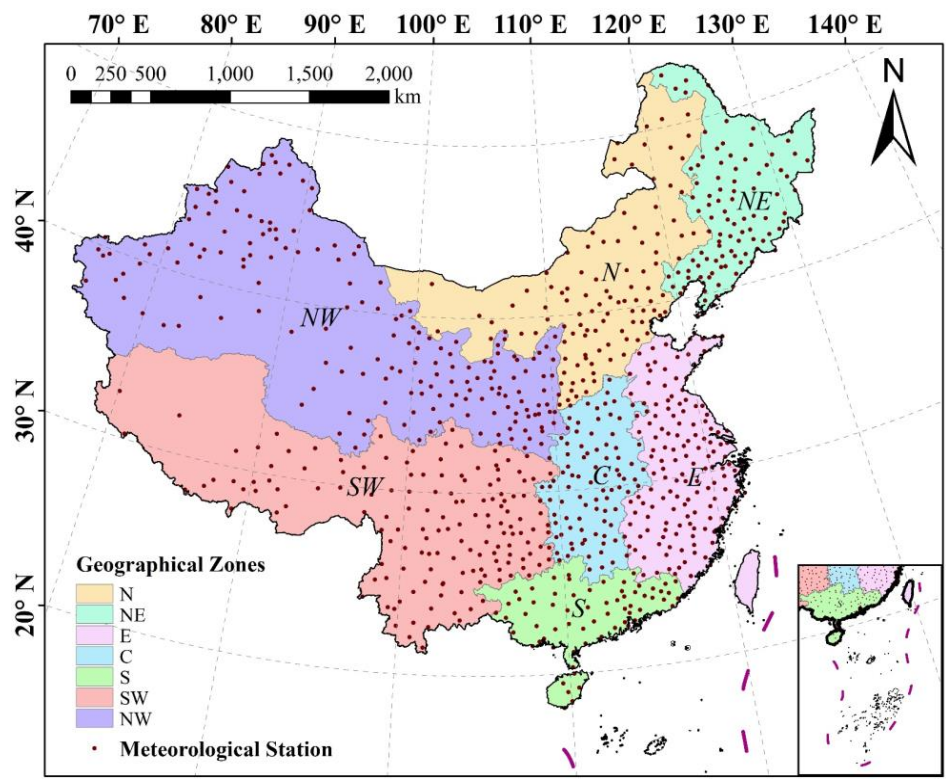

Fig. 1. Map of the study area and the geographical distribution of meteorological stations. N, E, S, NE, NW, SW

and C denotes northern China, eastern China, southern China, northeastern China, northwestern China, southwestern China and central China, respectively.

**2.2. Datasets**

**2.2.1. Meteorological Ta observation**

The meteorological Ta observations used in this study were obtained primarily from the China Meteorological Data Service Center (CMDC, http://data.cma.cn/). A total of 829 meteorological stations spread in mainland China were used. As shown in Fig. 1, there was a higher density of the stations located in southern, eastern and central China compared with the relatively sparse station distributions in the northwestern and southwestern areas, especially in Tibetan Plateau. Daily maximum Ta of these stations in 2015 was provided. Also, the accuracy and integrity of the observations have been improved after original calibration and quality control. In addition, the geographical and temporal parameters of each station provided simultaneously with the Ta observations were also used in this study, including latitude (Lat), longitude (Lon), month of year (Mon) and day of year (Doy).

**2.2.2. Remotely sensed data**

In this study, the remotely sensed datasets included LST, NDVI, land cover (Lc) and elevation (Ele).

**LST:** MODIS LST products have been proved to be an effective variable for estimating Ta in many previous studies (Chen et al., 2016; Noi et al., 2016; Zhang et al., 2011). Daily LST product, MOD11A1 (Terra Daily Land Surface Temperature and Emissivity) across China for the year 2015 were utilized in this study. MOD11A1 provided per-pixel temperature with a spatial resolution of 1 km sin grid, which was retrieved by the generalized split-window algorithm (Vancutsem et al., 2010). The accuracy of LST has been validated and reported to be better than 1 K under clear sky conditions in most cases (Wan et al., 2002; Wan, 2014). In this study, only daytime land surface temperature (LSTD) with the overpass time around 10:30 am local time was used. Considering that LSTD may be influenced by the observed angle, view zenith angle (Vangle) of day observation was also extracted from MOD11A1 product along with LSTD at 1 km spatial resolution.

**NDVI:** NDVI data was extracted from the MOD13A2 (Terra 16-Day Vegetation Index) product with a 1 km resolution. The 16-day NDVI product was used due to NDVI values do not change significantly within 16 days.

**Land cover:** Annual Terra and Aqua combined MODIS land cover product (MCD12Q1) for 2015 was used in this study. We reclassified land cover categories into six types across China including cropland, woodland, grassland, urban and built, water and barren in order to make it easier to assess the effect of land cover on the model performance. All the MODIS data used in this study were downloaded from the Level-1 and Atmosphere Archive and Distribution System Distributed Active Archive Center (LAADS DAAC, https://ladsweb.modaps.eosdis.nasa.gov/search/).

**Elevation:** Elevation data was obtained from the CGIAR Consortium for Spatial Information (CGIAR-CSI, http://srtm.csi.cgiar.org/index.asp), which provided global resampled Shuttle Radar Topography Mission (SRTM) digital elevation product with the spatial resolution of approximately 250 meters.

**2.2.3. Assimilation data**

The Global Land Data Assimilation System (GLDAS) makes full use of the advanced

generation of ground and space-based observation systems and provides a series of long-term gridded land surface states and flux parameters (Fang et al., 2009; Rodell et al., 2004). GLDAS version 2.1 datasets were downloaded from the NASA Goddard Earth Sciences Data and Information Services Center (GES DISC, https://disc.gsfc.nasa.gov/). Several assimilation data products such as wind speed (WS), soil moisture content at 0-10 cm underground (SoilMoi), albedo (Albe) and direct evaporation from bare soil (Esoil) with 0.25 degree and 3-hourly resolution were utilized, which were simulated with the Noah Land Surface Model 3.3 in Land Information System Version 7. The 3-hourly assimilation data were aggregated to a daily scale in this study.

**2.2.4. Socioeconomic data**

Socioeconomic factors were also taken into consideration to represent the influence of anthropogenic heat on Ta in a sense. Road density (RoadD) data was calculated within a 1° search radius by using the road network vector data downloaded from the OpenStreetMap (OSM, https://www.openstreetmap.org/). Only primary road, secondary road, tertiary road, trunk road, motorway and unclassified road were selected on account of the complexity of calculation. Besides, population density (PopD) data for 2015 was available on the Socioeconomic Data and Applications Center (SEDAC, http://sedac.ciesin.columbia.edu/). The original global raster data were produced with a 30 arc-second spatial resolution (CIESIN, 2017).

**2.2.5. Data pre-processing**

In total, we introduced station daily maximum Ta observation with some geographical and temporal parameters, remotely sensed data, assimilation data and socioeconomic data to conduct our work. All the data and their abbreviations were listed in Table 1. More specifically, the MODIS Reprojection Tool software was used to deal with the MODIS data from the original HDF-EOS format to GeoTIFF format. For assimilation data, the ArcPy site package of Python provided a productive way to process the original multidimensional netCDF file into a separate raster layer. For single raster data like elevation, population density and road density, pre-processing and processing are implemented by using ArcGIS software. After batch pre-processing of image mosaic, format conversion and image clip, remotely sensed data, assimilation data and socioeconomic data were reprojected to the same geographic coordinates system. For this work, nearest neighbor was chosen for resampling the raster data to the $0.01°×0.01°$ grid cell for consistency. Then, the nearest neighbor method was also used to match the point observations and raster data by extracting multiple corresponding variable values on the grid where each meteorological station was located. Considering that satellite-based data may be vacant or unusual due to the cloud cover or contamination and sensor fault (Shen et al., 2015; Wan, 2014; Zeng et al., 2018), eliminating unfilled and outlier data was necessary for the purpose of establishing effective data pairs. In total, 107578 matched samples with both daily maximum Ta and all predictor variable values were identified to form the experimental datasets. It should be noted that Ele and PopD value are highly differentiated in space, which may lead to unsatisfactory results. In order to narrow the range of these values, the original value is mapped by an exponential function in this study.

Table 1. Abbreviations of the data used in this study.

| Abbreviation | Data | Spatial resolution | Temporal resolution |
|---|---|---|---|
| Ta | Air temperature | - | 1 day |
| Lat | Latitude | - | - |
| Lon | Longitude | - | - |
| Doy | Day of year | - | - |
| Mon | Month of year | - | - |
| LSTD | Daytime land surface temperature | 1000 m | 1 day |
| NDVI | Normalized difference vegetation index | 1000 m | 16 days |
| Vangle | View zenith angle of day observation | 1000 m | 1 day |
| Ele | Elevation | 250 m | - |
| Lc | Land cover | 500 m | 1 year |
| PopD | Population density | 30" | 5 years |
| RoadD | Road density | Polyline | - |
| WS | Wind speed | 0.25° | 3 hours |
| SoilMoi | Soil moisture content | 0.25° | 3 hours |
| Albe | Albedo | 0.25° | 3 hours |
| Esoil | Direct evaporation from bare soil | 0.25° | 3 hours |

## 3. Methodology

### 3.1. Deep belief network

As the second generation of neural network, deep learning method was employed in this study to simulate the non-linear relationship between Ta observation and multi-source data. Compared with some general machine learning methods, deep learning makes it closer to artificial intelligence (Deng and Yu, 2014; Lecun et al., 2015). Deep belief network, as a Bayesian probability generation model, is developed for the first attempt to estimate Ta. The classic DBN can be regarded as a combination of multiple layers of simple, unsupervised networks, and usually is superposed by many restricted Boltzmann machines (RBM) layers and a back-propagation (BP) layer. Fig. 2 shows the structure of the DBN with two RBM layers as an example. Each RBM is a two-layers neural network, which consists of a visible input layer $V$ and a hidden layer $H$ (Hinton et al., 2006). As shown in Fig. 2, the training procedure of DBN can be treated as an efficient unsupervised layer-wise fashion (Hinton, 2009). Taking one RBM for example, supposing there are $n$ neurons in the visible layer and $m$ neurons in the hidden layer, the neurons have fully undirected connections between the two layers but no connections existing between neurons in the same layer. Generally, the parameters of the RBM such as weight matrix $W$, bias $a$ and $b$ are updated by the contrastive divergence (CD) algorithm (Hinton et al., 2006). The updating weight matrix $\Delta W_{ij}$ can be expressed as follows:

$$\Delta W_{ij} = \varepsilon \left( \langle V_i H_j \rangle_{data} - \langle V_i H_j \rangle_{model} \right) \tag{1}$$

where $V_i$ and $H_j$ are the states of the $i$th visible neuron and the $j$th hidden neuron, respectively; $\varepsilon$ is the learning rate; $\langle \cdot \rangle_{data}$ represents the expectation with respect to the distribution of the training samples; $\langle \cdot \rangle_{model}$ is the expectation under the partial derivative of the reconstructed model. Bias $a$ and $b$ are updated in a similar way.

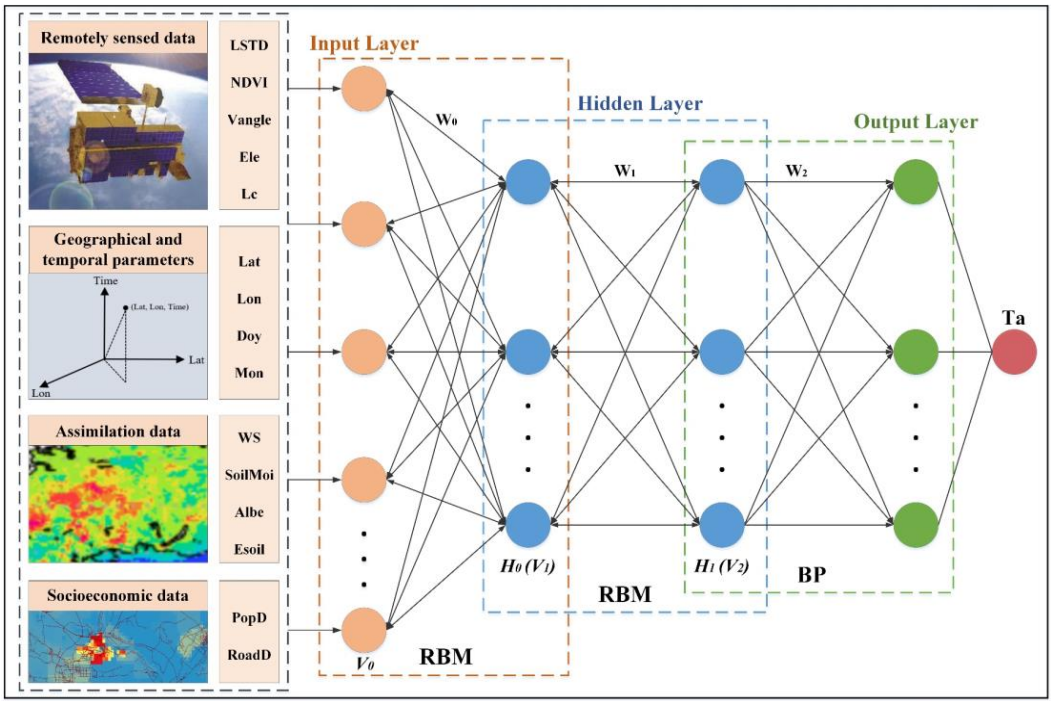

Fig. 2. The structure of the DBN model.

In this study, we utilized the DBN model by fusing Ta observations with geographical and temporal parameters, remotely sensed data, assimilation data and socioeconomic data. The structure of the model can be given by equation (2):

$$Ta = f(LSTD, NDVI, Vangle, Ele, Lc, Lat, Lon, Doy, Mon, WS, SoilMoi, Albe, Esoil, PopD, RoadD) \quad (2)$$

where $f()$ means the non-linear estimation function that the DBM model needs to model.

The flow chart of the DBN model used in this study is shown in Fig. 3, and the process can be summarized as pre-training, fine-tuning and prediction which is described in detail in the previous literature (Li et al., 2017). Once we put all the predictor variables into the model, they are recognized as the visible layer in the first RBM. The whole pre-training can be regarded as the process of extracting essential features from the input data that are associated with daily maximum Ta by using the feature optimization algorithm. This process is repeated from the lowest layer to the highest layer without supervision. Through the layer-by-layer pre-training until the hidden layer of the last RBM, the parameters can be obtained approximately close to their ideal values. In addition, we can obtain the estimated Ta, which are compared with the observed values. Then, the BP algorithm is adopted for fine-tuning all the weight parameters of the DBN model to get a refined prediction until the estimated error is small enough. Additionally, model cross-validation is implemented to evaluate whether the estimation results are accurate enough, otherwise we must constantly adjust the number of hidden layers in the model and the neurons in each hidden layer. Finally, a model with the best parameters can be applied to estimate spatially continuous Ta at the national scale. After many experiments, a 5-layers DBN model was developed in this study including one input layer, one output layer and three hidden layers. The number of neurons in each hidden layer is designed as 25, 20 and 15, respectively.

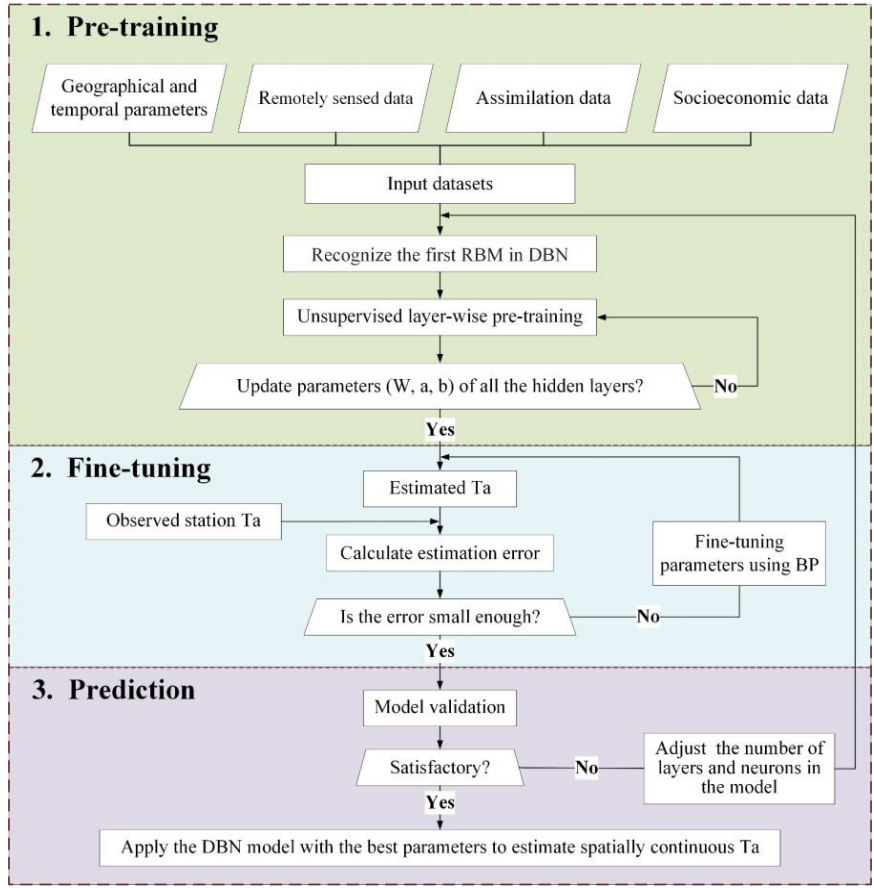

Fig. 3. The flow chart of the DBN model to estimate daily maximum Ta.

**3.2. Model validation**

To better evaluate the performance of each model, the ten-fold cross-validation approach was conducted to test the model predictive ability and overfitting problems (Rodríguez et al., 2010). In ten-fold cross-validation, all the samples were randomly split into ten groups of approximately equal size. Each group was withheld in turn as the validation dataset to assess the model performance, while the rest of the nine groups were used for model fitting. This procedure was repeated for ten times until each group had been tested exactly once as the validation part and got corresponding predictions. A set of statistical indicators-Pearson correlation coefficients (R), mean absolute error (MAE) and root mean square error (RMSE) were calculated to evaluate the model accuracy by comparing the estimated results with the real corresponding station-based observations.

## 4. Results

### 4.1. Descriptive statistics

Pearson correlation coefficients were calculated in this study to evaluate the strong linear or non-linear relationship between predictor variables and daily maximum Ta observations. The R values of all variables except Lc are presented in Fig. 4, since the Lc represents the categorical attributes rather than specific numerical meanings. It can be observed that LSTD has a strong correlation with Ta (R>0.9). NDVI, Albe, Esoil, SoilMoi, Lat, Ele, PopD and RoadD have a relatively moderate correlation with Ta with the absolute R values between 0.2 and 0.6. Those absolute R values below 0.2 indicate that WS, Vangle, Lon, Doy and Mon have an extremely weak correlation with Ta.

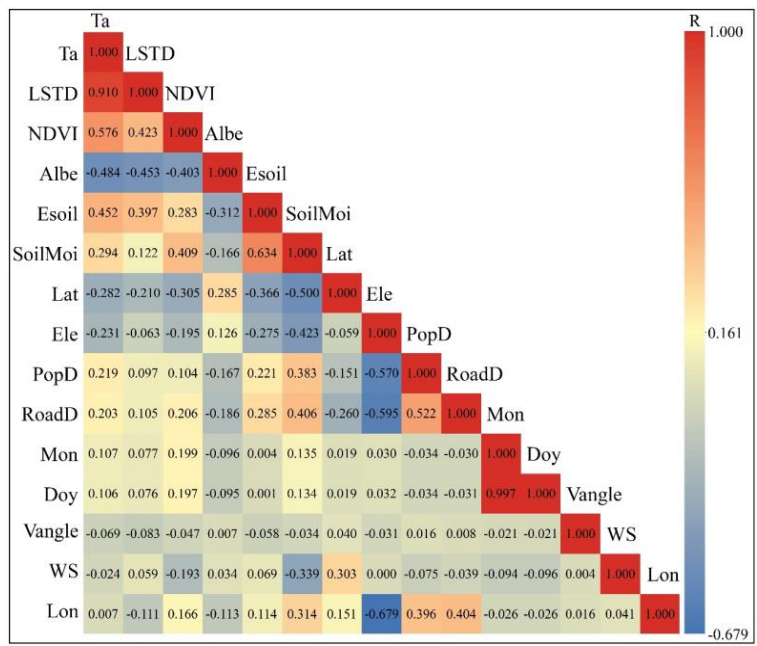

Fig. 4. Pearson correlation coefficient among the predictor variables and Ta.

Despite a strong positive correction between LSTD and daily maximum Ta is observed for the whole samples, the LSTD-Ta relationship is not constant under different circumstances.

To understand this phenomenon in detail, the correlation between LSTD and Ta was tested for different months, elevation ranges and latitude ranges to assess the spatial and temporal impact. From the results listed in Table 2, it is evident to note that the LSTD-Ta relationship is significantly influenced in diverse environments especially for different months with the distinct R value disparity reaches nearly 0.4 between May and January. As a whole, LSTD is more related to Ta in the areas with low elevation and high latitude as well as in the seasons with relatively low temperature. This phenomenon might lead to an obvious influence on the accuracy of Ta estimation.

Table 2. The R between LSTD and Ta for different months, elevation ranges and latitude ranges.

| Month / R | | | | Elevation (m) / R | | Latitude (°) / R | |
|---|---|---|---|---|---|---|---|
| Jan. | 0.926 | Jul. | 0.593 | <1000 | 0.954 | <20 | 0.803 |
| Feb. | 0.857 | Aug. | 0.549 | 1000-2000 | 0.918 | 20-30 | 0.800 |
| Mar. | 0.814 | Sep. | 0.638 | 2000-3000 | 0.855 | 30-40 | 0.880 |
| Apr. | 0.701 | Oct. | 0.799 | 3000-4000 | 0.761 | 40-50 | 0.955 |
| May. | 0.527 | Nov. | 0.899 | >4000 | 0.817 | >50 | 0.978 |
| Jun. | 0.541 | Dec. | 0.912 | | | | |

**4.2. Overall performance of the DBN model**

By applying all the above variables, daily maximum Ta across China was estimated and validated. Fig. 5 shows the overall performance for the DBN and other three contrast models. From the point density plots, the fitting line (the line in black) of the four models are nearly close to the 1:1 line (the line in red). However, it should be noted that MLR shows the most dispersed scatter dots, which indicates that using the linear model to derive Ta may cause relatively large errors in some conditions due to the complex relationships between LSTD and Ta. Unlike MLR, traditional BPNN is based on non-linear fitting, and the result is better than MLR but not ideal due to the relatively simple structure of the model. RF, as a new machine learning model, is also a non-linear technique which makes estimations by averaging an ensemble of individual regression trees, shows better results than MLR and BPNN. Especially, owing to the best capabilities of fitting non-linear relationships, deep learning method achieves the most accurate estimations with most of the scatter dots gathered close to the 1:1 line.

From another comprehensive and quantitative perspective, the cross-validation results provide a better understanding of the different performances for each model. Overall, all the models present acceptable fits with RMSE ranging from 1.996°C to 3.697°C, MAE ranging from 1.539°C to 2.879°C, and R ranging from 0.949 to 0.986, respectively, at the national scale. The best performance is achieved in deep learning method with the highest R value and lowest RMSE and MAE, followed by the RF and BPNN. Compared with the worst estimated result derived from the MLR, the RMSE decreases by 1.701°C (from 3.697 to 1.996), MAE decreases by 1.340°C (from 2.879 to 1.539) and R increases by 0.037 (from 0.949 to 0.986), which confirms that deep learning method has great potential capabilities in estimating Ta at the national scale. The reason may be that deep learning method can better simulate the potential characteristics of the variables, which are not intensively correlated with Ta than conventional methods like MLR, BPNN and RF.

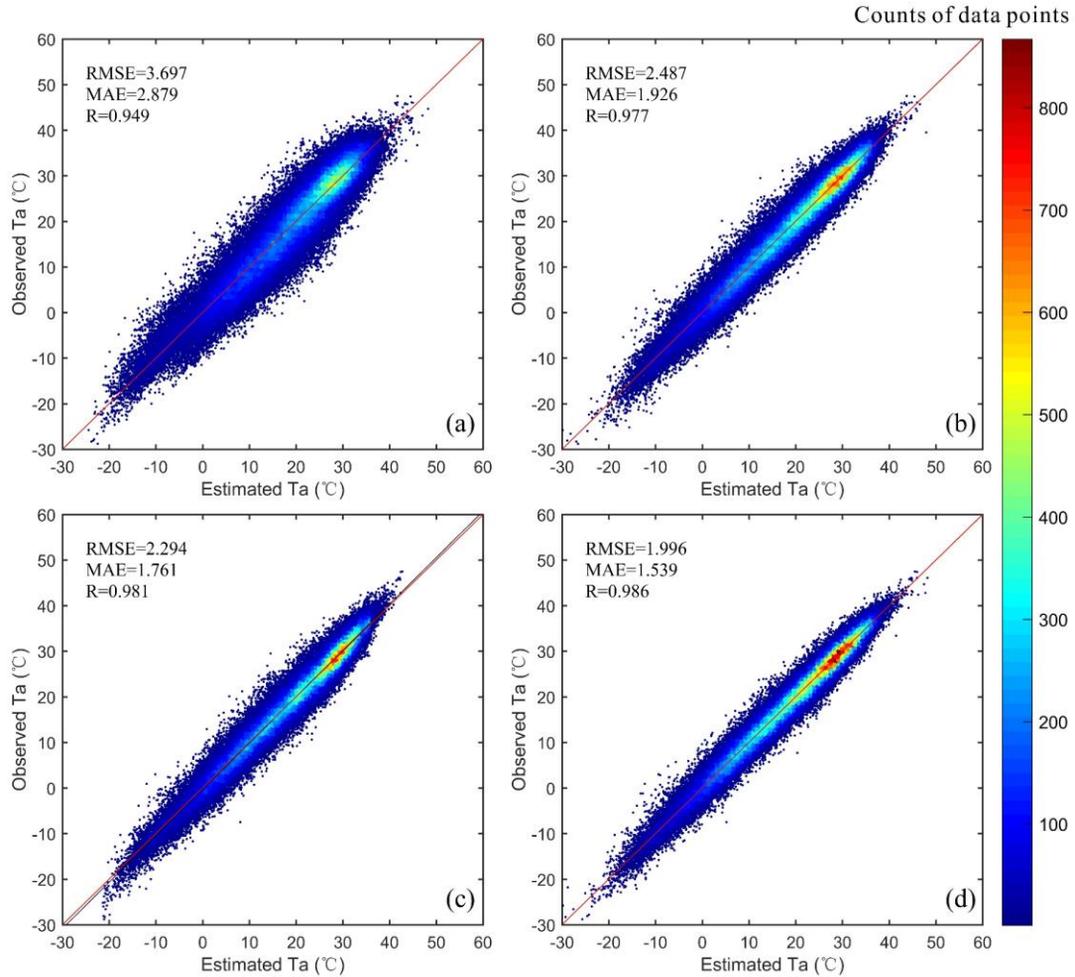

Fig. 5. The point density plots of the observed and estimated daily maximum Ta for the cross-validation results. (a)-(d) represent the results for MLR, BPNN, RF and DBN models, respectively. The line in black is the linear regression of the scattered dots; the line in red is the 1:1 line as reference.

**4.3. Spatial evaluation of model performance**

The model performance for the spatial pattern was evaluated, and the distribution of MAE for each meteorological station were shown in Fig. 6. There are significant variations existing in MAE spatial distribution. For MLR, the MAE range from 1.276 to 10.318°C with a huge fluctuation from one station to another, and the MAE of 86% of all stations are higher than 2°C (Fig. 6a). Most of the stations with high MAE are located in the southwestern and northwestern China. For BPNN, there are 30% of all stations report the MAE higher than 2°C and most of them are located in southwestern, northwestern and northern China (Fig. 6b).

For RF, the maximum MAE is reduced to 4.385°C with 17% of all stations are higher than 2°C. Spatially, the MAE exhibits relatively low values in southern and central China (Fig. 6c). Compared with the above three methods, the accuracy of the DBN model has been significantly improved with only 0.04% of all stations have relatively high estimation error (>2°C) in total (Fig. 6d). It is clear that the number of those stations with the MAE value more than 2.5°C is reduced to 5 for DBN, while there are 440, 37 and 23 stations for MLR, BPNN and RF, respectively. Additionally, the minimum and maximum MAE value for DBN are also reduced to about 0.742 and 4.289°C, respectively. Most of the MAEs change relatively gently with the values fluctuate steadily between 1 and 2°C for DBN. These results illustrate that deep learning method is more stable than conventional methods for most stations. In addition, from the overall distribution of MAEs in each model, stations located closer to the shoreline, where the climate is often dominated by the ocean, perform relatively poor. This is also consistent with some previous studies (Benali et al., 2012; Pelta and Chudnovsky, 2017). Besides, the stations located in southern China achieve better performance than northwestern stations in this study. This may benefit from the dense station distribution and the uncomplicated topographic and environmental conditions in southern China. In this study, the kernel density tool in ArcGIS was used to calculate the density for each station feature. Then, the relationship between the spatial density of stations and model performance was analyzed in detail. As shown in Fig. 7, the model performance in spatial is obviously influenced by the distribution of stations especially for MLR. Overall, there was a positive correlation between station density and model accuracy. That's to say, the more clustered the stations, the higher accuracy of the model. Compared with MLR, BPNN and RF, the lowest slope shown in Fig. 6d can reflect that DBN can reduce the effect of station density on accuracy to a certain extent.

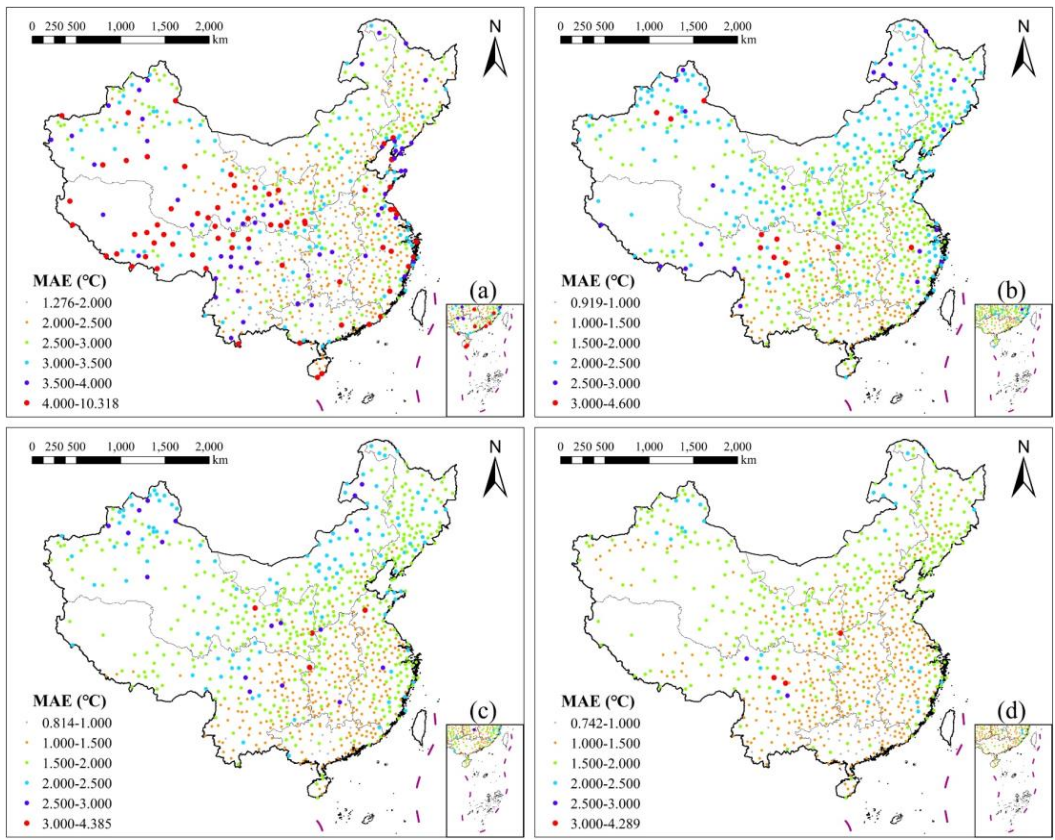

Fig. 6. The spatial distribution of MAE for each meteorological station. (a) MLR, (b) BPNN, (c) RF, (d) DBN.

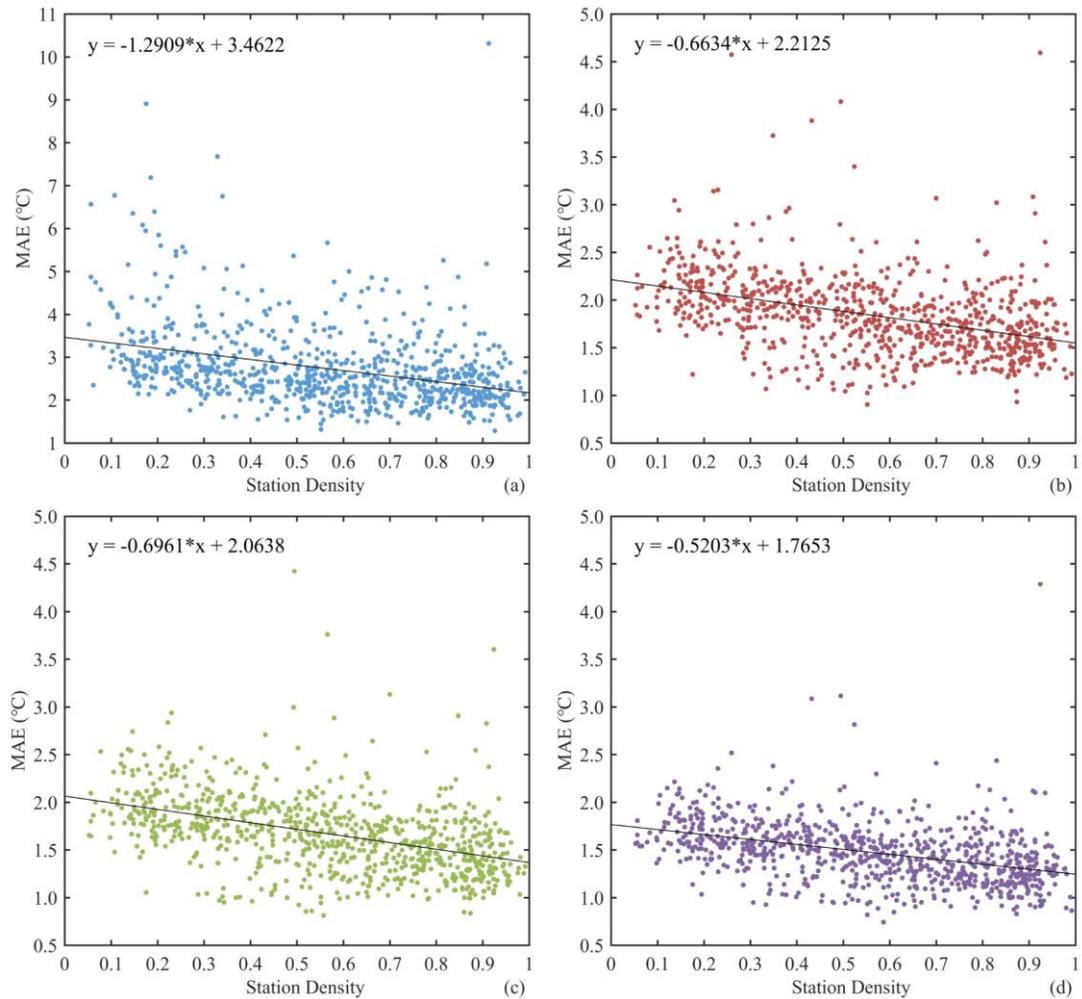

Fig. 7. The effect of station density on model performance. (a) MLR, (b) BPNN, (c) RF, (d) DBN.

Previous researches have revealed that land cover types have a significant influence on the relationship between LSTD and Ta (Cheng et al., 2008; Lin et al., 2012; Marzban et al., 2018). The model performance for different land cover types is compared by calculating the MAE. As shown in Fig. 8a, there is no doubt that the DBN model performed the best. However, the MAE varies at different land cover types for DBN, as well as in other models. The difference between the maximum and minimum MAE for DBN approximately reach to 0.325°C. In general, higher MAE values can be seen in woodland, barren and grassland. As for those stations located near the urban and built land, water and cropland, relatively better model performance is exhibited.

Apart from the above land cover types, model performance for different latitude ranges is also discussed in our study. The results shown in Fig. 8b suggest that DBN is superior to the other three models for all latitude ranges. For the DBN model, the MAE value varies from 1.331°C to 1.887°C. Overall, higher latitude may lead to poorer model performance except the latitude range from 20 to 30°. In this range, BPNN, RF and DBN all achieve the best model performance. For MLR, the model performance shows a wave-shaped curve for different latitude ranges.

In addition, it is widely acknowledged that Ta is highly related to elevation. In most cases, Ta may decrease as the elevation increases. Meanwhile, the correlation coefficient between LSTD and Ta shows obvious variations for different elevation ranges as mentioned in Section 4.1. Hence, the model performance for different elevation ranges is also analyzed in terms of MAE. From Fig. 8c, we can find out that models perform differently for specific elevation ranges, among which DBN model performs the best and shows the smallest MAE variation, followed by RF, BPNN and MLR in turn. Generally, the model performance for DBN is weakening by the elevation increasing. However, it is noteworthy to mention that there is a reverse when the elevation is above 4000 meters.

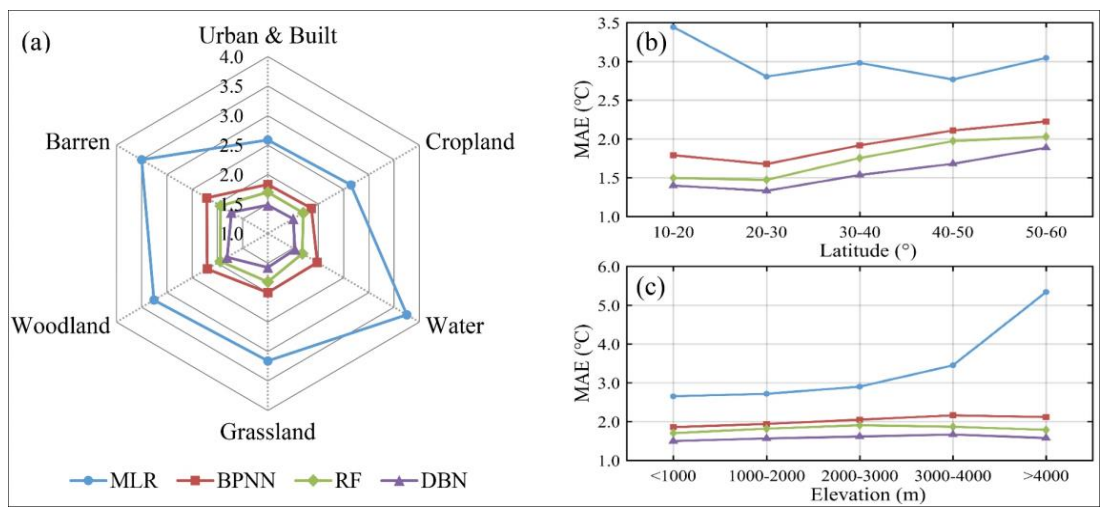

Fig. 8. Model performance for specific (a) land cover types, (b) latitude ranges and (c) elevation ranges.

**4.4. Temporal evaluation of model performance**

For temporal analysis, model performance at the monthly scale is evaluated. Fig. 9 shows the box plots of residual Ta values (estimated-observed) distribution for each month. It is clear that the medium of residuals for MLR is fluctuating. Besides, the medium of residual values from June to August are obviously below 0°C, which indicates that MLR tends to underestimate daily maximum Ta in these months. Compared with MLR, BPNN and RF methods, the medium of residuals for DBN all fall close to 0°C for each month, which indicates that deep learning method is not prone to cause overestimation or underestimation. Besides, DBN exhibits the lowest uncertainties in residual change than other models (Fig. 9d). In April, the residuals in BPNN and RF show obviously large fluctuations than the other months. Once we used deep learning method, this monthly variation can be reduced to a certain extent as shown in Fig. 9d.

From the MAEs for each model, we can see clearly that model uncertainties are rather high in term of month. Even for the DBN model, the difference between the maximum and minimum MAEs also reaches to about 0.494°C (Fig. 9d). The best model performance for the DBN model is achieved in August with the MAE of 1.248°C. More specifically, it's interesting to find that the DBN model seems to show the tendency to perform better in hot months, which is not consistent with our expectations. Because relatively low correlation coefficient between LSTD and Ta can be found in these months as demonstrated in Section 4.1. This finding reveals that model performance at specific circumstances may not strictly depend on the correlation coefficient between LSTD and Ta due to the efficient influence from other predictor variables. Overall, all the above-mentioned findings reinforce the necessity of considering temporal factors in Ta estimation models. To get a thorough understanding of this point, seasonal DBN models are worth attempting in the future work.

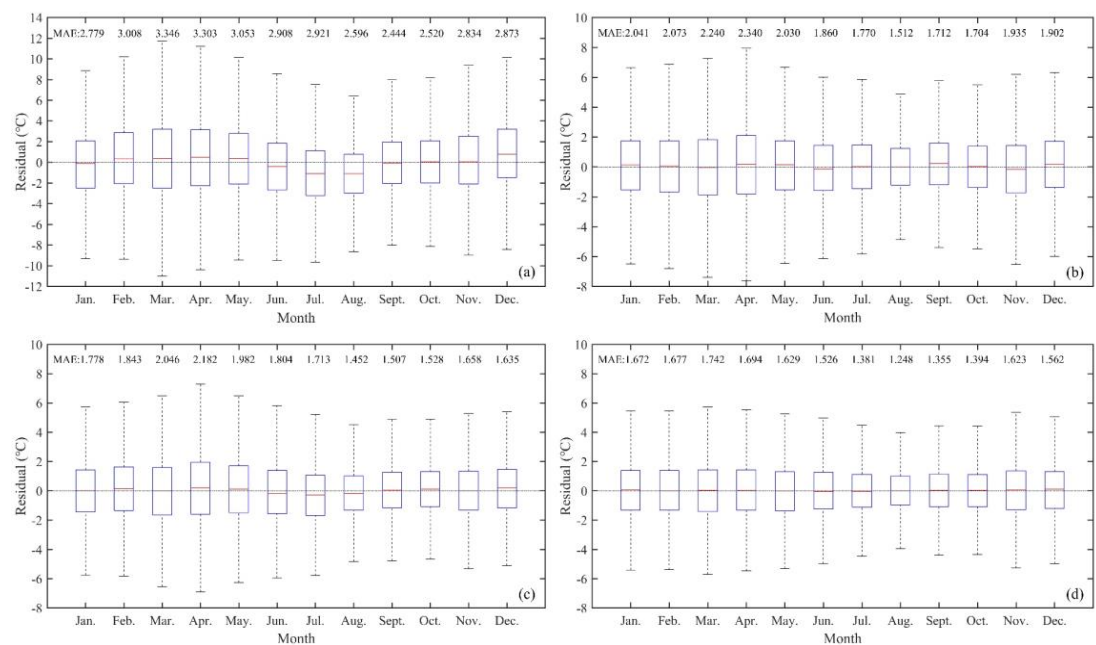

Fig. 9. Box plots of residuals for specific month. (a) MLR, (b) BPNN, (c) RF, (d) DBN.

## 5. Discussion

### 5.1. DBN structure comparison

To investigate the model performance, different adjustments of DBN structures were compared and the result was shown in the table below. With the increase of the number of DBN layers and neurons, the accuracy of the model increases obviously. However, when the hidden layer of DBN reaches to 4, the model performance tends to be stable. It's noteworthy that over-parameterized model may lead to overfitting. Besides, saving computational cost is another factor that we should take into consideration. Therefore, the DBN structure with three hidden layers, one input layer and one output layer is chosen in this study finally. The number of neurons in each hidden layer is designed as 25, 20 and 15, respectively.

Table 3. DBN model performance for different layers and neurons.

| DBN hidden layer | Hidden layer neurons | RMSE (°C) | MAE (°C) | R |
|---|---|---|---|---|
| 1 | 25 | 2.450 | 1.899 | 0.978 |
| 2 | 25 20 | 2.188 | 1.689 | 0.983 |
| 3 | 25 20 15 | 1.996 | 1.539 | 0.986 |
| 4 | 25 20 15 10 | 1.979 | 1.525 | 0.986 |

**5.2. DBN performance with different variables**

As stated, the DBN model is more promising than the other three conventional methods in Ta estimation. In addition, several types of datasets were fused in this study and some of the variables were even not linearly related to Ta as described in Section 4.1. To better understand the actual contribution of each dataset made to the DBN model, the model accuracy was evaluated with different combinations of the datasets. The statistical indicators for each combination are provided in Table 4. When only remotely sensed data are included in the DBN model, the RMSE, MAE and R is 3.099°C, 2.379°C, and 0.965, respectively. Compared with this basic estimated result, model accuracy is obviously improved when more variables are introduced in the model. For socioeconomic data and assimilation data, the MAE is decreased from 2.379°C to 2.118°C and 2.145°C, respectively. In addition, it should be noted that the geographical and temporal parameters were not strongly related to Ta but improved the model accuracy to a large extent. This phenomenon indicates that deep learning can effectively simulate the non-linear relationship between Ta and some predictor variables. Notably, the MAE of the combination R + S + A + P is decreased although it only changes a little compared with R + P, R + S + P and R + A + P. More importantly, the model accuracy for these different dataset combinations confirms that fusing multi-source data in the model actually make sense in Ta estimation. On the other hand, the mapping results will be optimized with more detailed information in spatial when more datasets are utilized.

Table 4. Model accuracy for different combinations of datasets.

| Datasets combinations | RMSE (°C) | MAE (°C) | R |
|---|---|---|---|
| R | 3.099 | 2.379 | 0.965 |
| R + S | 2.763 | 2.118 | 0.972 |
| R + A | 2.790 | 2.145 | 0.972 |
| R + P | 2.145 | 1.655 | 0.983 |
| R + S + A | 2.041 | 2.653 | 0.974 |
| R + S + P | 2.049 | 1.579 | 0.985 |
| R + A + P | 2.066 | 1.595 | 0.985 |
| R + S + A+ P | 1.995 | 1.539 | 0.986 |

R: Remotely sensed data; P: Geographical and temporal parameters; S: Socioeconomic data; A: Assimilation data.

## 5.3. Analysis of uncertainties in mapping results

Although deep learning showed superior overall and spatio-temporal model performance, and the variables introduced in the model truly improved the estimation accuracy, there are still two issues should be considered. The first point is that spatial over-fitting is a common problem in estimation researches, which means the model can perform well for the time series of stations, but fail in the estimation for some unknown locations, especially in some complicated and untrained area (Meyer et al., 2018, 2016). After many experiments, we found that the 16-days resolution NDVI data used in the model would lead to mapping outliers. The possible reason for this phenomenon may be that the model is unable to well fit the variable values with a regular temporal resolution of 16-days to the daily scale. The other point is that "nugget effects" phenomenon may be easily encountered when the variable value has a large variation in space or with regular spatial patterns (Molotch et al., 2005). To solve this problem, elevation and the population density data were processed by an exponential function as mentioned in Section 2.2.5. However, Lc data used in this study shows a regular distribution in space, which makes the results easily lead to "nugget effects" phenomenon. Considering that Lc does not represent a specific value, pre-processing the values seems to make no sense. Taking a local area for the 210th day as an example, the spatial distribution the Ta estimated by using all variables and the Ta estimated after removing the NDVI and Lc are presented in Fig. 10. As shown in Fig. 10a, "nugget effects" phenomenon can be obviously observed in the areas marked in the red circle. Additionally, some snow or ice-covered areas inside the red box shown in Fig. 10a present obvious higher estimations than surrounding areas, which is not consistent with the facts. Fortunately, after removing the NDVI and Lc, it can be seen clearly in Fig. 10b that these uncertainties in mapping results have been well addressed. More importantly, there was only a slight reduction in the model accuracy after removing these two variables, and the RMSE, MAE and R are 2.079°C, 1.606°C, 0.984, respectively. As a consequence, NDVI and Lc were removed as misleading variables for the DBN-based mapping of Ta across China.

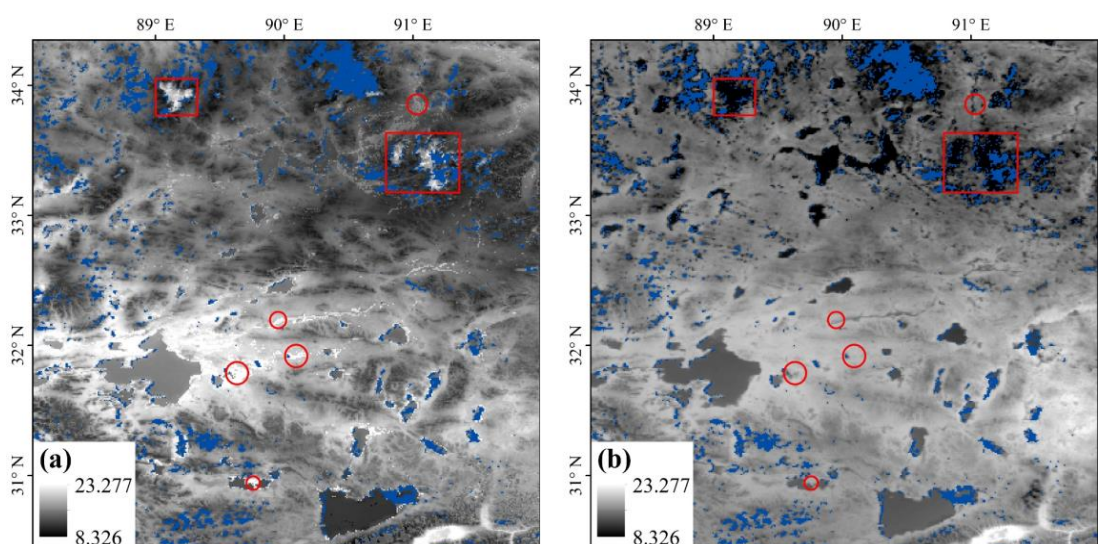

Fig. 10. Spatial distribution of (a) Ta estimated by using all variables and (b) Ta estimated after removing the NDVI and Lc over the local area for the 210th day (unit: °C). The blue pixels represent no data value.

### 5.4. Spatio-temporal distributions of Ta

After solving the over-fitting and "nugget effects" phenomenon in the mapping results, the 0.01° spatial resolution daily maximum Ta can be generated by the DBN model with accurate estimations. Taking the annual estimated Ta map as an example, we compared it with the corresponding assimilated Ta map provided by the GLDAS (Fig. 11). It should be noted that assimilated Ta were provided by the GLDAS as instantaneous variables with a 3-hourly resolution. Here, we calculated the maximum of eight Ta values per day as the daily maximum Ta value for each pixel. Additionally, assimilated Ta over water areas are not simulated by the GLDAS. In Fig. 11, the annual Ta for the DBN model shows a similar spatial distribution with the assimilated Ta, which suggest that DBN can well fit the general trend of Ta in space. Moreover, the DBN model exhibits more detailed spatial variations than the assimilated results, especially for some complex areas. For instance, the assimilated Ta in northwestern and southwestern China appears obvious pixel effect due to the low spatial resolution. These comparisons confirm the superiority of the DBN model and lend our confidence to advocate the DBN model in Ta estimation research in the future.

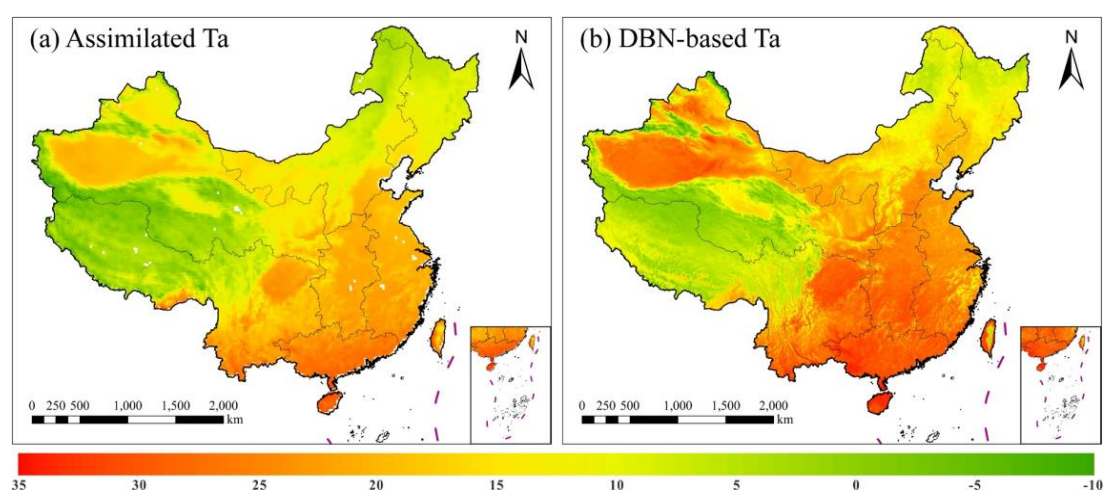

Fig. 11. Comparison of the annual daily maximum Ta map (unit: °C). (a) Assimilated Ta and (b) DBN-based Ta. The pixels in white represent no data values.

In addition, Ta mapping results for the four seasons are also presented in this study (Fig. 12). Seasonally, the continuous estimated Ta value ranges from -15 to 45°C in 2015. Ta values are obviously higher in summer (June to August) and lower in winter (December to February), while spring (March to May) and autumn (September to November) show the similar estimated mapping results. Spatially, it's clear that seasonal Ta generally decreases from the eastern and southern areas to the northeastern and southwestern areas, which is consistent with the physical situation. Ta values in summer are generally high, except for the high-elevation areas like the Tibetan plateau. In addition, water areas show lower Ta values than the surrounding land areas while the urban zones show relatively higher Ta than surrounding rural zones. These different trends may be caused by several factors, like topography condition, climate difference and social activity effects.

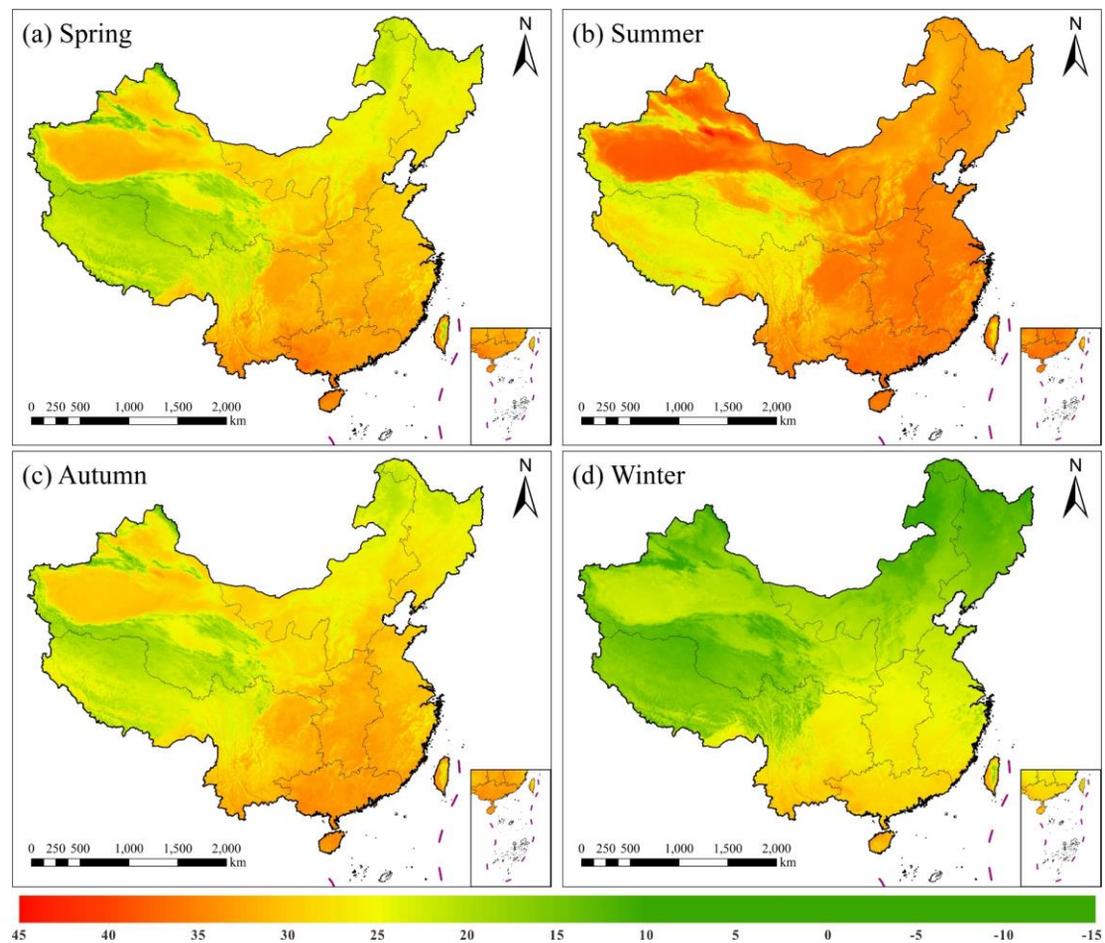

Fig. 12. The spatial patterns of seasonal Ta across China in 2015 estimated by the DBN model (unit: °C). (a) Spring, (b) Summer, (c) Autumn, (d) Winter. The pixels in white represent no data values.

## 6. Conclusions

In this study, the 5-layers DBN model, as a typical deep learning method, was employed to estimate Ta for the first attempt. Specific, we estimated the spatially continuous 0.01° daily maximum Ta across China by fusing remote sensing, station, simulation and socioeconomic data. Compared with conventional methods, the validation results showed that deep learning method could better take the non-linear relationship into consideration and achieved the best overall model performance with the RMSE of 1.996°C, MAE of 1.539°C, and R of 0.986. In addition, comprehensive analyses of the model performance for specific space and time were discussed, more accurate estimations could be obtained by using the DBN model. The performance of the DBN model with different combinations of datasets indicated that introduce effective variables in the model could improve the model performance to a great extent. Spatio-temporal Ta estimated by the DBN model showed more detailed spatial variations than assimilated Ta. Moreover, relevant researches can also be extended to the estimation of daily minimum and mean Ta in the future.

It must be emphasized that there are still several limitations in our study. On the one hand, although effective multi-source datasets can obviously improve the accuracy of the models, how to scientifically eliminate variables which are prone to lead uncertainties in mapping results is still a challenge. On the other hand, vacancy values reconstruction to improve the spatial coverage of LSTD data and quality assurance to ensure data availability need further exploration. Thus, future work should focus more attention on the missing information reconstruction of the estimated Ta caused by the incomplete LSTD. Finally, different models for specific regions and seasons at the national scale may worth further examination to improve the model performance.

**Acknowledgments**

This work was supported by the National Key R&D Program of China (no. 2016YFC0200900). We also acknowledge all the institutions for providing the data freely. The remotely sensed data are available from the Level-1 and Atmosphere Archive & Distribution System (https://ladsweb.modaps.eosdis.nasa.gov/). The station data are provided by the China Meteorological Data Service Center (http://data.cma.cn/). The assimilation data are obtained from the NASA Goddard Earth Sciences Data and Information Services Center (https://disc.gsfc.nasa.gov/). The socioeconomic data are downloaded from the OpenStreetMap (https://www.openstreetmap.org/) and the Socioeconomic Data and Applications Center (http://sedac.ciesin.columbia.edu/).

**References**

Benali, A., Carvalho, A.C., Nunes, J.P., Carvalhais, N., Santos, A., 2012. Estimating air surface temperature in Portugal using MODIS LST data. Remote Sens. Environ. 124, 108–121.

Chen, Y., Sun, H., Li, J., 2016. Estimating daily maximum air temperature with MODIS data and a daytime temperature variation model in Beijing urban area. Remote Sens. Lett. 7, 865–874.

Cheng, K.S., Su, Y.F., Kuo, F.T., Hung, W.C., Chiang, J.L., 2008. Assessing the effect of landcover on air temperature using remote sensing images-A pilot study in northern Taiwan. Landsc. Urban Plan. 85, 85–96.

CIESIN, 2017. Gridded Population of the World, Version 4 (GPWv4): Population Density, Revision 10.

Colombi, A., De Michele, C., Pepe, M., Rampini, A., Michele, C. De, 2007. Estimation of daily mean air temperature from MODIS LST in Alpine areas. EARSeL eProceedings 6, 38–46.

Deng, L., Yu, D., 2014. Deep learning: methods and applications. Found. Trends® Signal Process. 7, 197–387.

Fang, H., Beaudoing, H.K., Teng, W.L., Vollmer, B.E., 2009. Global Land data assimilation system (GLDAS) products, services and application from NASA hydrology data and information services center (HDISC).

Fu, G., Shen, Z., Zhang, X., Shi, P., Zhang, Y., Wu, J., 2011. Estimating air temperature of an alpine meadow on the Northern Tibetan Plateau using MODIS land surface temperature. Acta Ecol. Sin. 31, 8–13.

Hinton, G.E., 2009. Deep belief networks. Scholarpedia 4, 5947.

Hinton, G.E., Osindero, S., Teh, Y.-W., 2006. A fast learning algorithm for deep belief nets. Neural Comput. 18, 1527–1554.

Ho, H.C., Knudby, A., Sirovyak, P., Xu, Y., Hodul, M., Henderson, S.B., 2014. Mapping maximum urban air temperature on hot summer days. Remote Sens. Environ. 154, 38–45.

Hou, P., Chen, Y., Qiao, W., Cao, G., Jiang, W., Li, J., 2013. Near-surface air temperature retrieval from satellite images and influence by wetlands in urban region. Theor. Appl. Climatol. 111, 109–118.

J. Stoll, M., Brazel, A., 2013. Surface-air temperature relationships in the urban environment of Phoenix, Arizona. Phys. Geogr. 13, 160–179.

Janatian, N., Sadeghi, M., Sanaeinejad, S.H., Bakhshian, E., Farid, A., Hasheminia, S.M., Ghazanfari, S., 2017. A statistical framework for estimating air temperature using MODIS land surface temperature data. Int. J. Climatol. 37, 1181–1194.

Jin, M., Dickinson, R.E., 2010. Land surface skin temperature climatology: benefitting from the strengths of satellite observations. Environ. Res. Lett. 5.

Jin, M., Dickinson, R.E., Vogelmann, A.M., 1997. A comparison of CCM2–BATS skin temperature and surface-air temperature with satellite and surface observations. J. Clim. 10, 1505–1524.

Kuwata, K., Shibasaki, R., 2015. Estimating crop yields with deep learning and remotely sensed data, in: Geoscience and Remote Sensing Symposium (IGARSS), 2015 IEEE International. IEEE, pp. 858–861.

Lecun, Y., Bengio, Y., Hinton, G., 2015. Deep learning. Nature 521, 436–444.

Li, L., Lian, Z., Li, P., 2010. The effects of air temperature on office workers' well-being, workload and productivity-evaluated with subjective ratings. Appl. Ergon. 42, 29–36.

Li, L., Zha, Y., 2018. Mapping relative humidity, average and extreme temperature in hot summer over China. Sci. Total Environ. 615, 875–881.

Li, T., Shen, H., Yuan, Q., Zhang, X., Zhang, L., 2017. Estimating Ground-Level PM2.5 by Fusing Satellite and Station Observations: A Geo-Intelligent Deep Learning Approach. Geophys. Res. Lett. 44, 11985–11993.

Li, X., Zhou, Y., Asrar, G.R., Zhu, Z., 2018. Developing a 1 km resolution daily air temperature dataset for urban and surrounding areas in the conterminous United States. Remote Sens. Environ. 215, 74–84.

Lin, S., Moore, N.J., Messina, J.P., DeVisser, M.H., Wu, J., 2012. Evaluation of estimating daily maximum and minimum air temperature with MODIS data in east Africa. Int. J. Appl. Earth Obs. Geoinf. 18, 128–140.

Lowen, A.C., Mubareka, S., Steel, J., Palese, P., 2007. Influenza Virus Transmission Is Dependent on Relative Humidity and Temperature. PLOS Pathog. 3, 1–7.

Lu, N., Liang, S., Huang, G., Qin, J., Yao, L., Wang, D., Yang, K., 2018. Hierarchical Bayesian space-time estimation of monthly maximum and minimum surface air temperature. Remote Sens. Environ. 211, 48–58.

Marzban, F., Sodoudi, S., Preusker, R., 2018. The influence of land-cover type on the relationship between NDVI-LST and LST-Tair. Int. J. Remote Sens. 39, 1377–1398.

Meteotest, 2010. Meteonorm handbook, Part III: Theory Part 2. Accessed online in February 9 (2011).

Meyer, H., Katurji, M., Appelhans, T., Müller, M.U., Nauss, T., Roudier, P., Zawar-Reza, P., 2016. Mapping daily air temperature for Antarctica Based on MODIS LST. Remote Sens. 8, 1–16.

Meyer, H., Reudenbach, C., Hengl, T., Katurji, M., Nauss, T., 2018. Improving performance of spatio-temporal machine learning models using forward feature selection and target-oriented validation. Environ. Model. Softw. 101, 1–9.

Mohsenzadeh Karimi, S., Kisi, O., Porrajabali, M., Rouhani-Nia, F., Shiri, J., 2018. Evaluation of the support vector machine, random forest and geo-statistical methodologies for predicting long-term air temperature. ISH J. Hydraul. Eng. 1–11.

Molotch, N.P., Colee, M.T., Bales, R.C., Dozier, J., 2005. Estimating the spatial distribution of snow water equivalent in an alpine basin using binary regression tree models: the impact of digital elevation data and independent variable selection. Hydrol. Process. An Int. J. 19, 1459–1479.

Mostovoy, G. V., King, R.L., Reddy, K.R., Kakani, V.G., Filippova, M.G., 2006. Statistical Estimation of Daily Maximum and Minimum Air Temperatures from MODIS LST Data over the State of Mississippi. GIScience Remote Sens. 43, 78–110.

Nieto, H., Sandholt, I., Aguado, I., Chuvieco, E., Stisen, S., 2011. Air temperature estimation with MSG-SEVIRI data: Calibration and validation of the TVX algorithm for the Iberian Peninsula. Remote Sens. Environ. 115, 107–116.

Noi, P.T., Degener, J., Kappas, M., 2017. Comparison of Multiple Linear Regression, Cubist Regression, and Random Forest Algorithms to Estimate Daily Air Surface Temperature from Dynamic Combinations of MODIS LST Data. Remote Sens. 9, 398.

Noi, P.T., Kappas, M., Degener, J., 2016. Estimating daily maximum and minimum land air surface temperature using MODIS land surface temperature data and ground truth data in Northern Vietnam. Remote Sens. 8, 1002.

Pelta, R., Chudnovsky, A.A., 2017. Spatiotemporal estimation of air temperature patterns at the street level using high resolution satellite imagery. Sci. Total Environ. 579, 675–684.

Prihodko, L., Goward, S.N., 1997. Estimation of air temperature from remotely sensed surface observations. Remote Sens. Environ. 60, 335–346.

Robeson, S.M., 2002. Relationships between mean and standard deviation of air temperature: implications for global warming. Clim. Res. 22, 205–213.

Rodell, M., Houser, P.R., Jambor, U.E.A., Gottschalck, J., Mitchell, K., Meng, C.-J., Arsenault, K., Cosgrove, B., Radakovich, J., Bosilovich, M., 2004. The global land data assimilation system. Bull. Am. Meteorol. Soc. 85, 381–394.

Rodríguez, J.D., Pérez, A., Lozano, J.A., 2010. Sensitivity Analysis of k-Fold Cross Validation in Prediction Error Estimation. IEEE Trans. Pattern Anal. Mach. Intell. 32, 569–575.

Shen, H., Li, T., Yuan, Q., Zhang, L., 2018. Estimating regional ground-level PM2.5 directly from satellite top-of-atmosphere reflectance using deep belief networks. J. Geophys. Res. Atmos. 123, 875–886.

Shen, H., Li, X., Cheng, Q., Zeng, C., Yang, G., Li, H., Zhang, L., 2015. Missing Information Reconstruction of Remote Sensing Data: A Technical Review. IEEE Geosci. Remote Sens. Mag. 3, 61–85.

Shi, Y., Jiang, Z., Dong, L., Shen, S., 2017. Statistical estimation of high-resolution surface air temperature from MODIS over the Yangtze River Delta, China. J. Meteorol. Res. 31, 448–454.

Singh, R., Joshi, P.C., Kishtawal, C.M., 2006. A new method to determine near surface air temperature from satellite observations. Int. J. Remote Sens. 27, 2831–2846.

Song, X., Zhang, G., Liu, F., Li, D., Zhao, Y., Yang, J., 2016. Modeling spatio-temporal distribution of soil moisture by deep learning-based cellular automata model. J. Arid Land 8, 734–748.

Sun, Y.J., Wang, J.F., Zhang, R.H., Gillies, R.R., Xue, Y., Bo, Y.C., 2005. Air temperature retrieval from remote sensing data based on thermodynamics. Theor. Appl. Climatol. 80, 37–48.

Tomlinson, C.J., Chapman, L., Thornes, J.E., Baker, C.J., Prieto-Lopez, T., 2012. Comparing night-time satellite land surface temperature from MODIS and ground measured air temperature across a conurbation. Remote Sens. Lett. 3, 657–666.

Ung, A., Weber, C., Perron, G., Hirsch, J., Kleinpeter, J., Wald, L., Ranchin, T., 2001. Air pollution mapping over a city-virtual stations and morphological indicators. 10th Int. Symp. "Transport Air Pollution."

Vancutsem, C., Ceccato, P., Dinku, T., Connor, S.J., 2010. Evaluation of MODIS land surface temperature data to estimate air temperature in different ecosystems over Africa. Remote Sens. Environ. 114, 449–465.

Vogt, J. V, Viau, A.A., Paquet, F., 1997. Mapping Regional Air Temperature Fields Using Satellite-Derived Surface Skin Temperatures. Int. J. Climatol. 17, 1559–1579.

Wan, Z., 2014. New refinements and validation of the collection-6 MODIS land-surface temperature/emissivity product. Remote Sens. Environ. 140, 36–45.

Wan, Z., Zhang, Y., Zhang, Q., Li, Z. liang, 2002. Validation of the land-surface temperature products retrieved from terra moderate resolution imaging spectroradiometer data. Remote Sens. Environ. 83, 163–180.

Wang, L., Koike, T., Yang, K., Yeh, P.J.F., 2009. Assessment of a distributed biosphere hydrological model against streamflow and MODIS land surface temperature in the upper Tone River Basin. J. Hydrol. 377, 21–34.

Xu, Y., Knudby, A., Shen, Y., Liu, Y., 2018. Mapping Monthly Air Temperature in the Tibetan Plateau from MODIS Data Based on Machine Learning Methods. IEEE J. Sel. Top. Appl. Earth Obs. Remote Sens. 11, 345–354.

Xu, Y., Qin, Z., Yan, S., 2011. Estimation of near surface air temperature from MODIS data in the Yangtze River Delta. Trans. CSAE 27, 63–68.

Yao, Y., Zhang, B., 2013. MODIS-based estimation of air temperature and heating-up effect of the Tibetan Plateau. Acta Geogr. Sin. 68, 95–107.

Zakšek, K., Schroedter-Homscheidt, M., 2009. Parameterization of air temperature in high temporal and spatial resolution from a combination of the SEVIRI and MODIS instruments. ISPRS J. Photogramm. Remote Sens. 64, 414–421.

Zeng, C., Long, D., Shen, H., Wu, P., Cui, Y., Hong, Y., 2018. A two-step framework for reconstructing remotely sensed land surface temperatures contaminated by cloud. ISPRS J. Photogramm. Remote Sens. 141, 30–45.

Zhang, L., Huang, J., Wang, X., 2014. A Review on Air Temperature Estimation by Satellite Thermal Infrared Remote Sensing. J. Nat. Resour. 29, 540–552.

Zhang, R., Rong, Y., Tian, J., Su, H., Li, Z., Liu, S., 2015. A Remote Sensing Method for Estimating Surface Air Temperature and Surface Vapor Pressure on a Regional Scale. Remote Sens. 7, 6005–6025.

Zhang, W., Huang, Y., Yu, Y., Sun, W., 2011. Empirical models for estimating daily maximum, minimum and mean air temperatures with MODIS land surface temperatures. Int. J. Remote Sens. 32, 9415–9440.

Zhu, S., Zhang, G., 2011. Progress in Near Surface Air Temperature Retrieved by Remote Sensing Technology. Adv. Earth Sci. 26, 724–730.

Zhu, W., Lű, A., Jia, S., 2013. Estimation of daily maximum and minimum air temperature using MODIS land surface temperature products. Remote Sens. Environ. 130, 62–73.

Zhu, W., Lű, A., Jia, S., Yan, J., Mahmood, R., 2017. Retrievals of all-weather daytime air temperature from MODIS products. Remote Sens. Environ. 189, 152–163.